\documentstyle[12pt,aaspp4]{article}

\lefthead{Metzger \& Ma}
\righthead{Abell 697}
\def\lta{\mathrel{\rlap{\lower 3pt\hbox{$\mathchar"218$}}
     \raise 2.0pt\hbox{$\mathchar"13C$}}}
\def\gta{\mathrel{\rlap{\lower 3pt\hbox{$\mathchar"218$}}
     \raise 2.0pt\hbox{$\mathchar"13E$}}}
\def\kms{km~s$^{-1}$}

\def\etal{{\it et~al.}} 
\def\pf{\ifmmode{{\hbox{\sc psf}}}\else{{\sc psf}}\fi}
\def\mM{\ifmmode(m{-}M)\else$(m{-}M)$\fi}
\def\msun{\ifmmode{\hbox{M$_\odot$}}\else{M$_\odot$}\fi}

\begin{document}
\tighten

\title{Lensed Arcs and Inner Structure of Abell~697\altaffilmark{1}}

\author{Mark R. Metzger \altaffilmark{2} and
Chung-Pei Ma \altaffilmark{3}}
\altaffiltext{1}{Based on observations obtained at the 
W.M. Keck Observatory, operated as a scientific partnership
by the California Institute of Technology, the University of California,
and the National Aeronautics and Space Administration.
}
\altaffiltext{2}{Division of Physics, Mathematics, and
Astronomy, California Institute of Technology, MS 105-24,
Pasadena,~CA~91125; {\tt mrm@caltech.edu}}

\altaffiltext{3}{Department of Physics and Astronomy,
University of Pennsylvania, Philadelphia, PA 19104; 
{\tt cpma@strad.physics.upenn.edu}}

\begin{abstract}

We present new optical observations of the $z=0.282$ cluster Abell 697
from the Keck II telescope.  Images show an unusual disturbed
structure in the cD halo and a previously unknown faint gravitational
lens arc.  A spectrum of the arc did not yield a redshift, but its
spectrum and colors suggest it lies at $z>1.3$.  We construct models
to reproduce the arc that show the potential is likely to be highly
elliptical.  We suggest that this cluster may have undergone a recent
merger and is in the process of forming its cD galaxy.  Analysis of
X-ray data from ROSAT and ASCA suggests that the merging process is
sufficiently advanced that the gas in the cluster has relaxed, and
A697 lies near the $L_x$-$T_x$ relation for normal clusters.

\end{abstract}


\section{Introduction}

Clusters of galaxies represent the largest scales in the universe at
which gravitational collapse has become non-linear and virialization
takes place.  As such, they also are the warehouses of most of the
visible baryons in the universe: hydrogen gas that falls into the
cluster and heats in the process, forming an intracluster medium too
hot to be bound by individual galaxies.  Massive clusters therefore
also tend to be strong sources of X-ray emission via thermal
bremsstrahlung of the intracluster gas, which provides a means for
clusters to be identified in addition to the overdensity of visible
galaxies.

Clusters also serve as cosmological probes, as their formation and
evolution rate depend on cosmological parameters and the kinematics of
the dark matter.  While optical surveys have led to the discovery of
most clusters at low redshift, most of the known high-redshift
clusters have been discovered through X-ray surveys (e.g. the EMSS
survey, \cite{Gio90}).  To interpret cluster evolution from high
redshift to the present requires a proper understanding of the
differences between X-ray and optical cluster samples, each with their
advantages and pitfalls.
Abell 697 (at $z=0.282$) is a cluster that represents an overlap
between the two samples.  Discovered optically (\cite{ACO89}), it is fairly
rich and is one of the higher redshift clusters in the ACO catalog.
It is also X-ray luminous and has sufficient flux to be included in
X-ray selected cluster samples (\cite{Ebe98}).

We report new Keck imaging and spectroscopic observations of the
central 6\arcmin\ of A697.  The images show a highly
asymmetric cD halo around A697, with low surface brightness components
that may be remnants of recent merger activity.  Our imaging also
reveals a faint, blue, arc-like feature which is likely to be lensed
by the cluster core.  Deep long-slit spectroscopy of the arc did not
yield a direct redshift for this object, but does provide several
cluster galaxy redshifts to estimate a velocity dispersion for the
cluster.  The combination of the blue color of the arc and the lack of
strong emission lines in the optical spectrum suggest that the arc is
at $z \gta 1.3$.  We construct a simple model of the cluster potential
near the core to explain the lensed feature, and suggest two
possibilities for the more extended, arc-like red structure found to
the northwest of the core.

\section{Observations and Data Reduction}

\begin{figure}
	\vspace{7in}
	\includegraphics{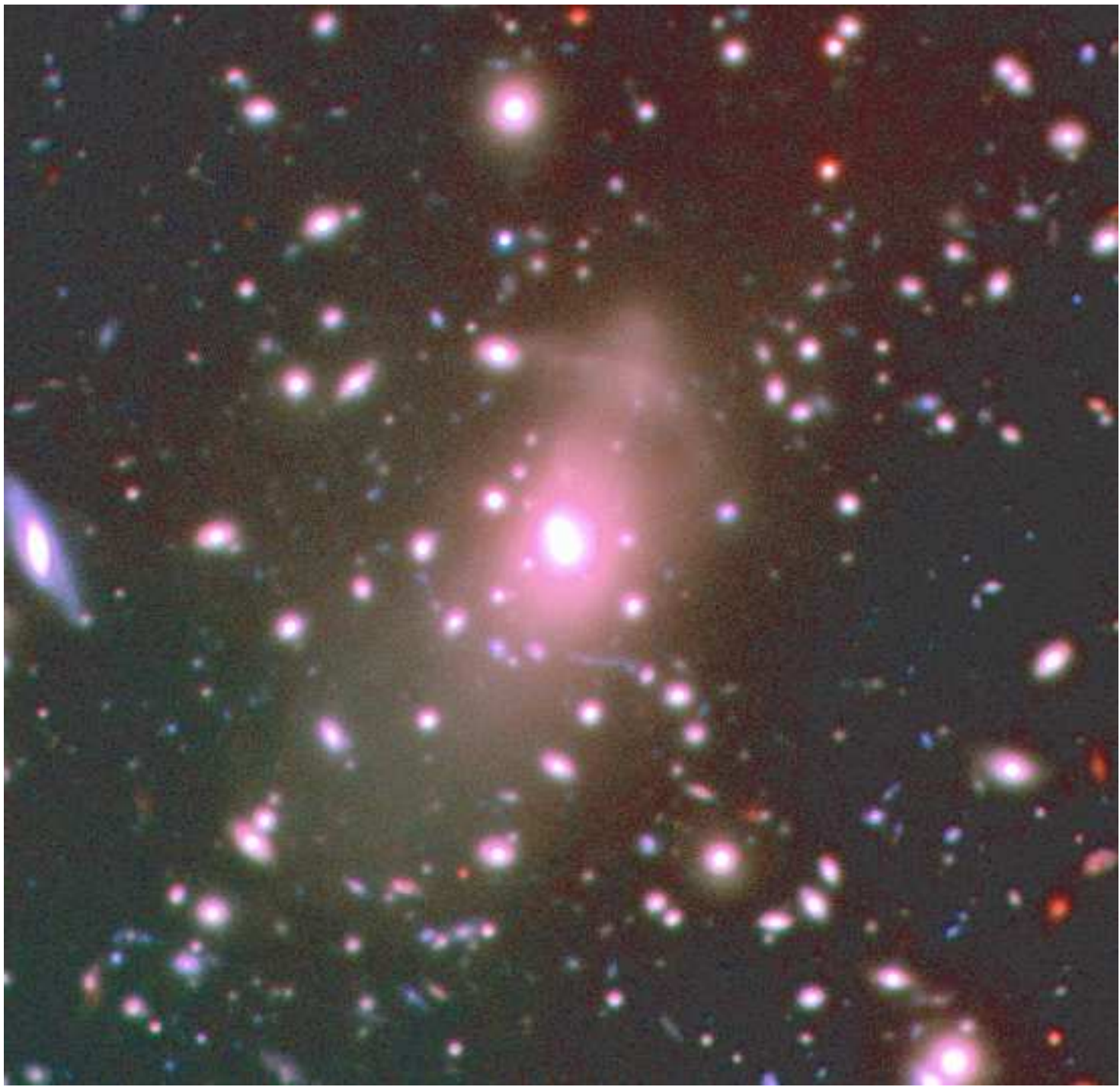}
\caption{
A false-color Keck/LRIS image of the central 120\arcsec\ of Abell 697.  Blue,
green, red correspond to the V, R, I band images, respectively.
North is up and east is to the left.  The lensed arc is
visible to the south of the cluster, and the extended red source to
the northwest.  Exposure times were 3000 seconds in each band, and
reaches a surface brightness limit of 28.2 mag arcsec$^{-2}$ in V.}
\label{fig:img}
\end{figure}

Images of the central region of A697 were obtained on UT January
28, 1998 on the Keck II 10 m telescope with LRIS (\cite{Oke95}).  The
optics illuminate a $6\arcmin\times 8\arcmin$ field onto a 2048 $\times$
2048 CCD with pixels projecting to $0\farcs211$ square on the sky.
Multiple dithered images were obtained in each of the $V$, $R$, and
$I$ bands, which were debiased, flatfield-corrected
with twilight sky illumination, and corrected for gain variation.
The images were then registered and combined using the
``Mora'' reduction package (\cite{Met94}); bad pixels 
were removed from the stack, and the final image was constructed using a
weighted sum.  The total exposure time was 3000 s in $V$ and
$R$, and 2700 s in $I$.  Photometric calibration was provided by
observing standard fields of Landolt (1992)\nocite{arlo}.
Figure~\ref{fig:img} shows a color image of the inner 4\arcmin\ of the
cluster.

A quick reduction of the images at the telescope revealed a blue
arc-like structure to the southwest of the cD.  We proceeded to obtain
three dithered 1200\,s longslit spectra of the arc with LRIS using a
300 $\ell$ mm$^{-1}$ grating on UT January 30, 1999, at a sky position
angle of 83$.^\circ$8.  The individual frames were processed with Mora
to remove the amplifier bias and nonlinearity, and to correct the
response with internal halogen lamp spectra.  The three frames were
aligned and co-added, rejecting cosmic rays using an iterative filter.
Wavelength calibration was obtained using Hg-Kr-Ne-Ar lamp spectra,
and the arc spectrum was extracted from the 2-d frame by adding flux
over the region where the arc was visible on the image and subtracting
sky from adjacent uncontaminated regions on the slit.  Flux
calibration was obtained using a spectrum of Feige 67 (\cite{Oke90}).
No significant emission or absorption lines are visible in the
arc spectrum.  When averaged over 500 \AA\ bins, we find only a
marginal continuum detection from the arc, but do not have enough
signal-to-noise to obtain an absorption redshift via cross-correlation or
template matching.

\begin{figure}
	\vspace{4.5in}
	\includegraphics{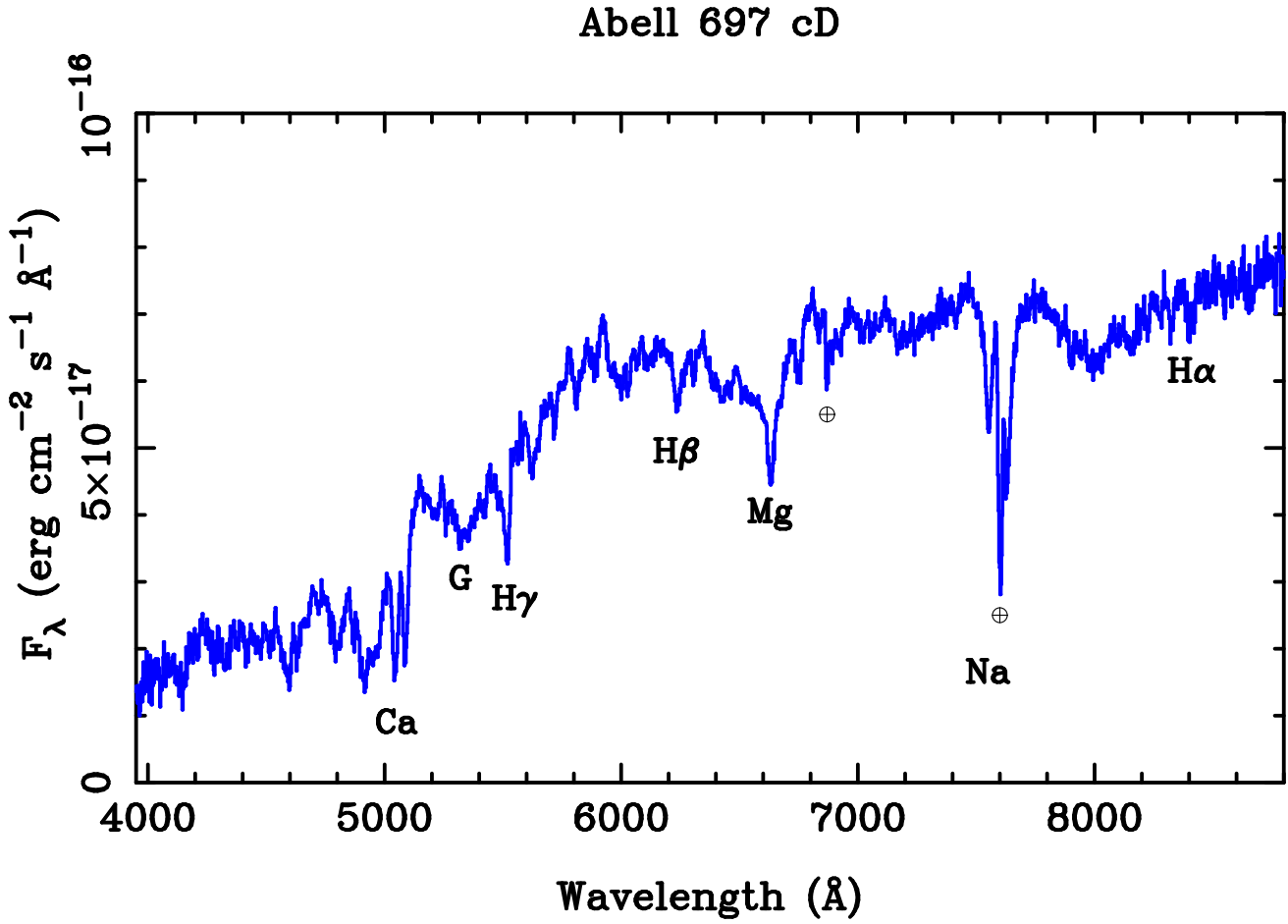}
\caption{A spectrum of the Abell 697 cD galaxy core, obtained with Keck/LRIS
using a 1\farcs0 slit.  Telluric O$_2$ A- and B-band absorption features are
marked.  No significant H$\alpha$ emission is detected.}
\label{fig:spec}
\end{figure}

An additional 1200\,s spectrum was taken centered on the bright cD
nucleus (see Fig.~\ref{fig:spec}) and aligned with the secondary
nucleus at a sky PA of 28$^\circ$.  The two long-slit spectra also
capture a number of nearby galaxies in the field,
which are included in our sample to estimate of the cluster velocity
dispersion.  For the galaxies without strong emission lines, the
well-exposed spectrum of the main cD nucleus was used as a
cross-correlation template; we measure $z=0.2824$ from the \ion{Ca}{2}
and H$\beta$ absorption features, in good agreement with the value
measured by \cite{Cra95}.  The velocities for 11 galaxies relative to
the cD are given in Table 1; 9 of these can be considered members of
A697 based on proximity in redshift space.  Using all 9 galaxies, we
derive a velocity dispersion of $\sigma = 941 \pm 296$ \kms\ (rest
frame).  However, the distribution of the velocities is non-Gaussian:
two of the galaxies lie at relatively large velocities from the main
cD nucleus, at $-2735$ \kms\ and $-2044$ \kms.  Removing these reduces
$\sigma$ to 553 \kms\ from 7 galaxies, at a mean velocity relative to
the cD of $93 \pm 85$ \kms.  The possibility of the presence of
subgroups dictates that more velocities are needed to obtain a
reliable dispersion, though our two estimates probably give reasonable
bounds.  The lower dispersion, however, would be unusually low for a
cluster with such a high intracluster gas temperature and X-ray
luminosity (see \S~\ref{sec:clus}).

We also attempted to measure dispersion within the cD halo itself as a
function of radius, but significant contamination from interloping
galaxies prevented an accurate estimate by artificially enhancing the
apparent line widths.  Our cD spectrum also shows no sign of H$\alpha$
emission to an upper limit of $W_\lambda({\rm obs}) \leq 0.5$\AA, in
contrast to the Crawford \etal\ (1995) measurement of $3.2 \pm 1.4$\AA.

\section{Cluster Properties}\label{sec:clus}

The central region of the cluster is dominated by a bright elliptical
galaxy with a cD halo profile (Fig.~\ref{fig:img}), having total V
magnitude of $18.1$.  A secondary nucleus is also present 2\arcsec\ to
the northeast.  At a radius of 5\arcsec, the cD halo has a roughly
elliptical profile with axis ratio $b/a = 0.78$, but at larger radius
the halo becomes significantly asymmetric with a large extension
visible to $r=40\arcsec$ to the southeast at surface brightness of V=28
mag arcsec$^{-2}$.  The orientation of the inner cD isophotes, at
roughly PA$=163^\circ$, is well aligned with the ASCA 2--10 keV image
brightness contours at larger radius (see below).

An unusual, low surface brightness extended feature is seen to the
northwest of the cD center.  The surface brightness varies irregularly
from 24.5 to 25.5 V mag arcsec$^{-2}$ over the central region covering
about 100 arcsec$^2$.  The color of much of the extended source is
consistent with that of an old stellar population at this redshift (as
well as the cluster ellipticals), suggesting it is comprised of stars
and is possibly a remnant of tidally disrupted galaxies.  Part of the
source extends in the direction of one of the cluster ellipticals, but
superposed on this wide extension is a more narrow, bluer arc-like
structure that may be an image of a lensed background galaxy (see
\S~\ref{sec:model}).

X-ray observations of A697 with ROSAT and ASCA were obtained from the
NASA HEASARC public data archive.  A697 appears near the edge of the
field of a 10800 sec ROSAT PSPC observation of GBS 0839+37, and is
cataloged in the ROSAT Bright Cluster Survey (\cite{Ebe98}) with total
luminosity of 1.6 $\times$ 10$^{45}\,h_{50}^{-2}$ erg s$^{-1}$ in the
energy range from 0.1--2 keV (observed frame).  The data gives a wide
field of view to the north and west of the cluster center, showing no
other hot clusters nearby that might be part of an aligned large-scale
structure.  A 28000 HRI image centered on the field likewise shows no
bright extended structures to the southeast.  The ASCA data from both
GIS and SIS were reduced and analyzed using the XSPEC software package
(\cite{Arn96}).  A fit to the spectrum using a Raymond-Smith model as
implemented in XSPEC yields a temperature of $10.2 \pm 1.9$ keV, which
lies fairly close to the Lx-Tx relation (Mushotzky \& Scharf 1998).
The data also reveal an extended x-ray source to the southwest (off of
our optical images), possibly associated with an additional infalling
group.  Ellipses fit to the X-ray isophotes at 6\arcmin\ major axis
diameter give an axis ratio of $b/a = 0.68$ oriented at ${\rm
PA}=155^\circ$.

\section{Lens Models}\label{sec:model}

We use an elliptical potential of the form
$$
\psi(r^\prime) = 4 \pi {\left( {\sigma_{1D} \over c } \right)}^2
{{D_{LS}} \over {D_S}} ( r_c^2 +  {r^\prime}^2 )^{1 \over 2} \,
$$
to model the cluster as a lens (\cite{Bla87}).  The variable
$\sigma_{1D}$ is the line-of-sight velocity dispersion in the limit
$r^\prime \gg r_c$ for the spherical case, $D_S$ and $D_{LS}$ are
angular size distances to the source and from the lens to the source,
respectively, and $r_c$ defines a softening radius.  The ellipticity
enters in the definition of $r^\prime$, ${r^\prime}^2 = (1 -
\epsilon_p) x^2 + (1 + \epsilon_p) y^2 \,,$ where the coordinates
$x,y$ are aligned along the major and minor axes of the potential (see
\cite{MFK93} for details about $\epsilon_p$) .  This represents a
softened isothermal sphere for $\epsilon_p = 0$, and models the shape
of dark halo potentials fairly well near the cores of clusters (e.g.,
Tyson \etal\ 1998).  We choose the more limited relation with a fixed
power law, as there are few lensing constraints in this system to
differentiate various potential profiles.

We search for the best-fit model by minimizing the distances between
the predicted image positions and the observed arc positions with
respect to the free model parameters.  In each iteration, the observed
arc pixels are first mapped onto the source plane using the current
model parameters.  We then find all the image locations of each source
with a grid searching algorithm that maps a fine grid in the lens
plane to the source plane and then scans for pixels in the lens plane
whose mapped pixels coincide with the source positions.

\begin{figure}
	\vspace{4.5in}
	\includegraphics{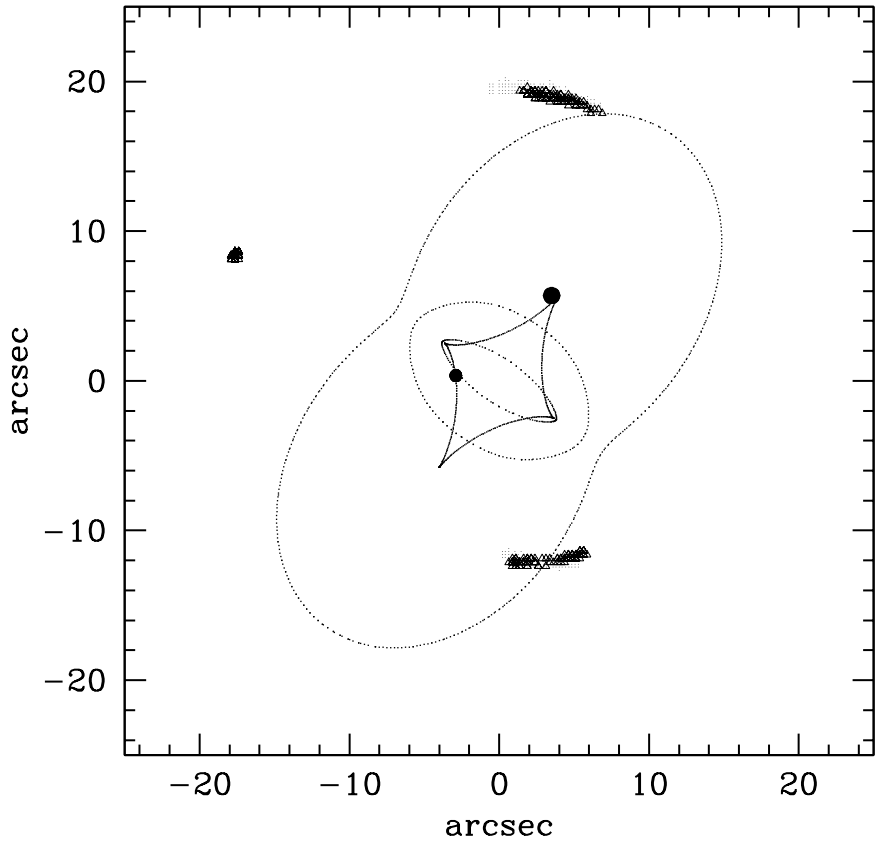}
\caption{
A plausible model for the Abell 697 arc
candidates.  The filled circle centered at (-2.9, 0.35) indicates the
source position (assuming $z_s=1.4$) for the lower arc.  The predicted
counter image is at about (-17, 8).  The inner two contours are the
caustics and the outer two are the critical curves.  The filled circle
centered at (3.5, 5.7) labels the source position (with $z_s=1.2$) for
the upper arc candidate.}
\label{fig:model}
\end{figure}

Although few observational constraints are available in this system,
the unusual orientation of the southern arc-like object makes this
lens nontrivial to model.  Its center of curvature points about
5\arcsec\ away (toward the west) from the cD center, and this feature
cannot be modeled by a single potential centered at the cD with the
measured orientation $\theta=73^\circ$ (i.e. PA$=163^\circ$) and
potential ellipticity $\epsilon_p=0.08$ that is suggested by the cD
profile.  Given the disturbed surface brightness distribution
surrounding the cD, however, it is conceivable that the shape of the
cluster potential does not trace the cD halo exactly.  Once $\theta$
and $\epsilon_p$ for the cluster are allowed to vary, we are able to
construct a simple model that reasonably reproduces the position and
shape of the southern arc (see Fig.~\ref{fig:model}).  The potential
for the cluster in this model is centered at the cD and has a velocity
dispersion $\sigma_{1D}=910$ km~s$^{-1}$, core radius $r_c=10\arcsec$,
ellipticity $\epsilon_p=0.22$, and orientation $\theta=55^\circ$.  The
source is placed at $z_s=1.4$, consistent with the lower limit of 1.3
from our spectrum.  An additional singular elliptical potential with
$\sigma_{1D}=200$ km s$^{-1}$ and the measured cD profile is used to
represent the cD galaxy.  As Figure~\ref{fig:model} shows, the
southern arc is a fold arc, and the counter-image appears east of the
cD.  Our ground-based image shows a number of faint blue, arclike
candidates in this area.  High resolution HST data will be needed for
more detailed morphological information and more refined modeling.
Due to the lack of arclike objects in the area for possible 4th and
5th images, we favor a three image lens model (here two merge to form
an arc).  The ellipticity of 0.22 helps to flatten the contour of the
radial caustic shown in Figure~\ref{fig:model} and allows the source
to be placed outside of it while remaining near a fold of the
tangential caustic.

The northern arc-like object can also be reproduced in our lens model.
The lack of a counterarc suggests that the simplest way to accommodate
this candidate arc is for it to be a single distorted image of a red
galaxy, which can occur if the source is outside of both caustics and
the arc is outside of both critical curves.  This scenario would
require the source redshift to be below 1.4, given that the
tangential critical curve for the source of the southern arc at
$z_s=1.4$ is already nearly touching the northern arc.
Figure~\ref{fig:model} illustrates a plausible source position for the
northern object assuming $z_s=1.2$.

\section{Discussion and Summary}\label{sec:disc}

Together, the asymmetric cD profile, extended features, and highly
elliptical potential implied by the arc model suggest that A697 may
have recently undergone a significant merger event (e.g. Roettiger et
al. 1996).  In this scenario, the ellipticity and more highly extended
outer contours are a dynamical consequence of the merger, in which the
tidal tails produced in the encounters have not had time to relax.  A
more highly elliptical potential could result from two massive dark
matter cores that have not completed the merger.  Several
observational consequences to the intracluster gas would also be
predicted in this case, including regions of non-uniform, elevated
temperature and X-ray luminosity (\cite{Evr96}; \cite{Ric98}; Jones et
al. 1997; \cite{Chu99}).

Another scenario that could explain the complex structure is that the
cD is undergoing the process of forming its extended halo, possibly
from the merging of several massive galaxies.  The dark matter
simulations made by Dubinski (1998) suggest that this scenario can
produce massive galaxies similar in appearance to a cD.  His figure
3 shows a cD in the process of formation, and several features
including tidal stripping and halo asymmetry qualitatively agree with
our observations.  However, many of the galaxies near the cluster core
appear tidally disrupted, while few of the concentrated ellipticals
show this effect in A697.  Given that these simulations did not
include intracluster gas and were of finite resolution, though, it is
difficult to draw any quantitative conclusions (though it is
suggestive).  The significant disruption of the cD halo and the presence
of the low surface brightness source, interpreted as a formation event,
provides some evidence in favor of the major-merger scenario for cD halo
formation over more gentle processes (e.g. \cite{Lop97}), but of course may
not be the only formation mechanism.

The detailed X-ray structure of A697 will be an interesting probe into
the physics of this merger.  Simple models may not fit, as recent
mergers may leave shocks in gas producing a complex inner temperature
and brightness structure.  More detailed models may help to confirm
a recent merger, as some of these structures should dissipate rapidly,
and specific emission lines from the X-ray gas should be measurable
with new instruments aboard XMM and Chandra.  Such a model of the
merging inner potential would also benefit from deep HST images of the
core, from which additional faint arcs or arclets could provide further
model constraints and resolution.


Our deep optical imaging of A697 reveals several arc-like objects
and a complex inner structure.  A spectrum of the lower arc did not
display obvious emission lines; combined with its blue color, this
indicates $z>1.3$ for the lensed galaxy.  We have
constructed a potential model to reproduce the arc's tilted
orientation.  The model agrees to within the rather large
uncertainties of the velocity dispersion that was estimated from nine
galaxy spectra.  A possible arc on the northern side of the cluster
can also be reproduced in this model.  The cD itself, while having a
normal spectrum, has an unusual and highly disturbed halo profile,
including possible remnants of recently disrupted old cluster
galaxies.  The potential model implied by the arc structure is also
highly elliptical, and we suggest that this is due to a recent
merger event, either between subcluster elements or massive galaxies
acting to form the cD.  We also suggest that this cluster would
provide an ideal system for testing numerical simulation predictions
of the effects of cluster merging on the intracluster gas
with high-resolution images and spectra in optical and X-ray
wavelengths.

\acknowledgments

We thank Roger Blandford, Una Hwang, and John Blakeslee for helpful
discussions, and Terry Stickel and Ron Quick for assistance at the
telescope.  This research has made use of NED, operated by the Jet
Propulsion Laboratory, Caltech, under contract with NASA, and data
obtained from the HEASARC online service, provided by NASA's Goddard Space
Flight Center.  M.R.M. acknowledges support from Caltech, NASA grant
AR-07520.01-96A, and NSF grant AST 9820664.  C.-P. M. acknowledges
support of an Alfred P. Sloan Foundation Fellowship, a Cottrell
Scholars Award from the Research Corporation, a Penn Research
Foundation Award, and NSF grant AST 9973461.





\clearpage

\begin{deluxetable}{ccc}
\newdimen\digitwidth
\newdimen\minuswidth
\setbox0=\hbox{\rm0}
\digitwidth=\wd0
\catcode`?=\active
\def?{\kern\digitwidth}
\setbox0=\hbox{\rm--}
\minuswidth=\wd0
\catcode`!=\active
\def!{\kern\minuswidth}
\tablecaption{Galaxy redshifts, relative to Abell 697 cD at z=0.2824
\label{tab:gals}}
\tablewidth{300pt}
\tablehead{
\colhead{Galaxy ID} &
\colhead{$cz$ (\kms)}	& 
\colhead{$\sigma(cz)$}
}
\startdata
cD & !???0 & ??0 \nl
1 & !?714 & 120 \nl
3 & !?278 & ?70 \nl
4 & !??25 & 161 \nl
6 & -2044 & 321 \nl
7 & !?758 & ?30 \nl
11 & !??14 & 314 \nl
14 & -2735 & 304 \nl
20 & ?-779 & 155 \nl
21 & ?-358 & 284 \nl
\enddata
\end{deluxetable}

\end{document}